\documentclass{PoS}

\usepackage{xspace}

\usepackage{soul}

\def\aSC{{\kappa_{\rm sc}}}

\newcommand{\pythia}{{\sc Pythia}\xspace}

\title{Understanding wide jet suppression in data through the hybrid strong/weak coupling model}

\ShortTitle{Understanding wide jet suppression in data through the hybrid model}

\author{ \speaker{Daniel Pablos},$^{ab}$ Jorge Casalderrey-Solana,$^{cd}$ Guilherme Milhano$^{ef}$ and Krishna Rajagopal$^{g}$\\
     \llap{$^a$}Department of Physics and Astronomy, Wayne State University, Detroit MI 48201\\
     \llap{$^b$}Department of Physics and Astronomy, McGill University, Montr\'{e}al QC H3A-2T8\\
     \llap{$^c$}Rudolf Peierls Centre for Theoretical Physics, University of Oxford, Clarendon Laboratory, Parks Road, Oxford OX1 3PU United Kingdom\\
     \llap{$^d$}Departament de F\'\i sica Qu\`antica i Astrof\'\i sica \& Institut de Ci\`encies del Cosmos (ICC), Universitat de Barcelona, Mart\'{\i}  i Franqu\`es 1, 08028 Barcelona, Spain\\
     \llap{$^e$}LIP, Av. Prof. Gama Pinto, 2, P-1649-003 Lisboa , Portugal\\
     \llap{$^f$}Instituto Superior T\'ecnico (IST), Universidade de Lisboa, Av. Rovisco Pais 1, 1049-001, Lisbon, Portugal\\
     \llap{$^g$}Center for Theoretical Physics, Massachusetts Institute of Technology, Cambridge, MA 02139 USA\\
     E-mail:  \email{pablosd@physics.mcgill.ca}, \email{jorge.casalderrey@ub.edu}, \email{gmilhano@lip.pt}, \email{krishna@mit.edu}}

\abstract{We explore a set of jet substructure observables that use grooming techniques such as the Soft Drop procedure by performing simulations with the hybrid strong/weak coupling model for jet quenching. The results obtained for the observables presented in this proceedings, namely the number of Soft Drop splittings, or $n_{\rm SD}$, and the sharing momentum distribution $z_g$ for different angular cuts between the two main branches in a jet, can be easily understood as arising from the fact that among all the jets with a given jet $p_T$, those jets which are wider in opening angle, meaning that they have a higher jet mass and come from showers featuring
more splittings, tend to lose more energy than the narrower jets. We comment on the comparison to data from ALICE and CMS, and point out the caveats arising from the consideration of 
smearing effects 
due to the presence of a large fluctuating background.}

\FullConference{International Conference on Hard and Electromagnetic Probes of High-Energy Nuclear Collisions\\
		30 September - 5 October 2018\\
		Aix-Les-Bains, Savoie, France}

\begin{document}


The study of the modification of high energy jets traversing the deconfined QCD matter created in heavy ion collisions represents one of the best tools to access the properties of quark-gluon plasma (QGP). This new state of matter, whose bulk flow is well described by the hydrodynamic evolution of a strongly coupled droplet of QGP liquid, interacts strongly with the partons originating from the hard scattering, making them lose energy to the plasma and modifying the properties of the final reconstructed jets. In this work, we study a new generation of jet substructure observables with very promising features, exploring how these observables are modified in jets that have traversed a droplet of QGP.
%
%
%
We focus in particular on the subset of groomed observables proposed by ALICE \cite{nima}, which extend the pioneering analysis done by CMS \cite{dhanush} and provide crucial new inputs. Grooming techniques \cite{Larkoski:2014wba} have recently gained a lot of attention in the field of heavy ions because, for jets produced in proton-proton collisions, they give access to core features of a parton shower such as the shape of the splitting function by relating it to the measurable momentum sharing distribution $z_g$. This motivates studying the potential modifications to the $z_g$ distribution caused by passage through the medium.
By performing simulations within the hybrid strong/weak coupling model 
\cite{Casalderrey-Solana:2014bpa}, 
we will show how the physics that dominates the relative behavior of hadron and jet suppression explained in Ref.~\cite{Casalderrey-Solana:2018wrw} can also help to understand 
this set of preliminary results.




\section{The hybrid strong/weak coupling model}
The hybrid strong/weak coupling model relies on the scale separation between the high virtuality $Q$ inherited from the hard scattering, responsible for the parton shower evolution, and the temperature $T$ that governs the dynamics of the QGP as well as the dominant physics of the jet/QGP interaction. In this way, we can assume that the virtuality relaxation process through successive splittings is governed by the perturbative DGLAP evolution equations, since $Q \gg T$, while the cross-talk between the shower partons and the medium is assumed to be described by nonperturbative physics, since $T \sim \Lambda_{\rm QCD}$. In order to implement these ideas, we rely on the event generator \pythia to produce the hard scatterings and the subsequent shower evolution. We assign a lifetime to the different partons following a formation time argument so that we can keep track of the space-time evolution of the shower. In between splittings, we extract energy and momentum from the partons according to a differential energy loss rate, obtained through holographic techniques for hot $\mathcal N=4$ SYM theory in the limit of infinite coupling and $N_c$ \cite{Chesler:2014jva}, 
which reads:

\begin{equation}\label{CR_rate}
\left. \frac{d E_{\rm parton}}{dx}\right|_{\rm strongly~coupled}= - \frac{4}{\pi} E_{\rm in} \frac{x^2}{x_{\rm therm}^2} \frac{1}{\sqrt{x_{\rm therm}^2-x^2}} \quad , ~~~~~~~~~  x_{\rm therm}= \frac{1}{2\aSC}\frac{E_{\rm in}^{1/3}}{T^{4/3}}
\end{equation}
where $E_{\rm in}$ is the parton's initial energy, $x_{\rm therm}$ is the stopping distance and $\aSC$ is an $\mathcal{O}(1)$ parameter that depends on the details of the particular gauge theory. The final hadrons with which we perform our analysis come from both the surviving shower partons, which fragment through the Lund string model present in \pythia, and the soft hadrons coming from the decay of the wake generated by the injection of energy and momentum into the plasma, which we estimate by solving the Cooper-Frye spectrum assuming small perturbations and instantaneous hydrodynamization \cite{Casalderrey-Solana:2014bpa}.

\section{Results}
In this Section we provide results from the hybrid model without finite resolution effects (as described in Ref.~\cite{Hulcher:2017cpt}; to be studied in future work), using the 
range of values for $\aSC$ obtained by performing a global fit to central hadron and jet data in Ref.~\cite{Casalderrey-Solana:2018wrw}. We compute the jet substructure observables by implementing the grooming techniques \cite{Larkoski:2014wba} implemented in FastJet \cite{Cacciari:2011ma}. All cuts and jet selections are chosen according to those adopted by ALICE in their preliminary results~\cite{nima}, against which we will quantitatively confront the results from our model in future work. The parameters of the Soft Drop procedure are $z_{\rm cut}=0.1$ and $\beta=0$. This setting removes large angle soft radiation. What remains from 
the medium response is then only its contribution near the core of the jet which is almost negligible  for our treatment of the wake in the plasma. (This need not be so in other treatments; see, e.g., the sizable effect shown for the notably harder JEWEL recoils.)

\begin{figure}[t!]
  \centering
  \includegraphics[width=.36\textwidth,height=.28\textwidth]{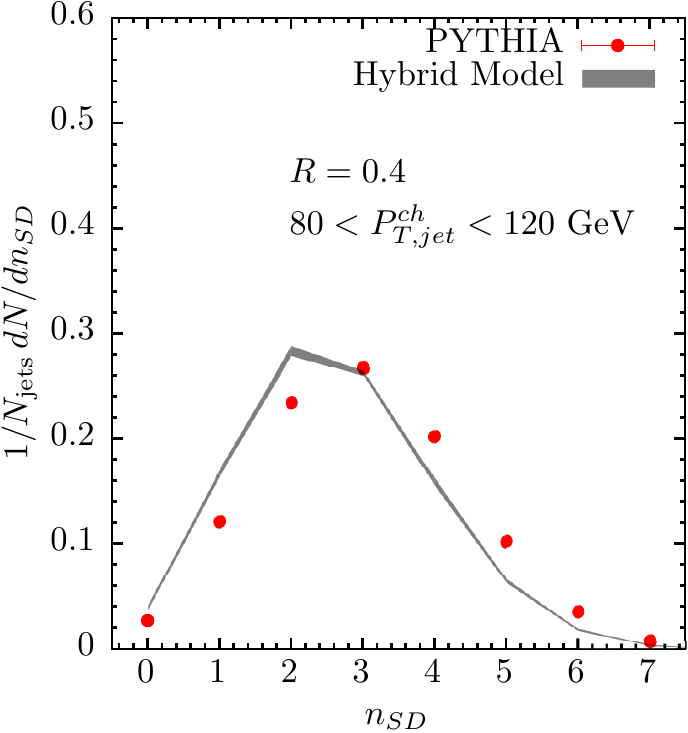}
    \vspace{-0.1in}
  \caption{Distribution of the number of Soft Drop splittings, $n_{\rm SD}$, normalized to the total number of jets. The entry at $n_{\rm SD}=0$ corresponds to jets that never satisfy the Soft Drop condition. The vacuum results from \pythia calculations (red dots) are compared to the thin grey band representing hybrid model results.}
  \label{nSD}
\end{figure}

The curves in Fig.~\ref{nSD} show 
the distribution of the number of times emissions from a single jet's hardest branch satisfy the Soft Drop condition, referred to as $n_{\rm SD}$.  
Jets with larger $n_{\rm SD}$ are those whose showers feature more splittings. 
We compare results from hybrid model calculations for heavy ion collisions to vacuum (\pythia) results. 
One can observe a clear, although mild, medium modification: a small reduction in $n_{\rm SD}$. 
Within the hybrid model, this effect can be attributed to the fact that those jets that split more (the wider ones) are quenched more, on average, than those jets that split less (the narrower ones), meaning that the latter have a larger representation in the final jet sample after quenching than they 
would have had without quenching.
Indeed, the total amount of energy lost by a given jet reconstructed with a given anti-$k_t$ radius parameter $R$ increases with the number of independent energy loss sources (e.g. partons) 
resolved by the medium, provided that the energy and momentum transferred to the medium efficiently ends up at angles greater than $R$. The results shown in the next figure further support this interpretation.


\begin{figure}[t]

	\centering 
	
	\vspace{-0.1in}
	\hspace*{\fill}
	\includegraphics[scale=0.55]{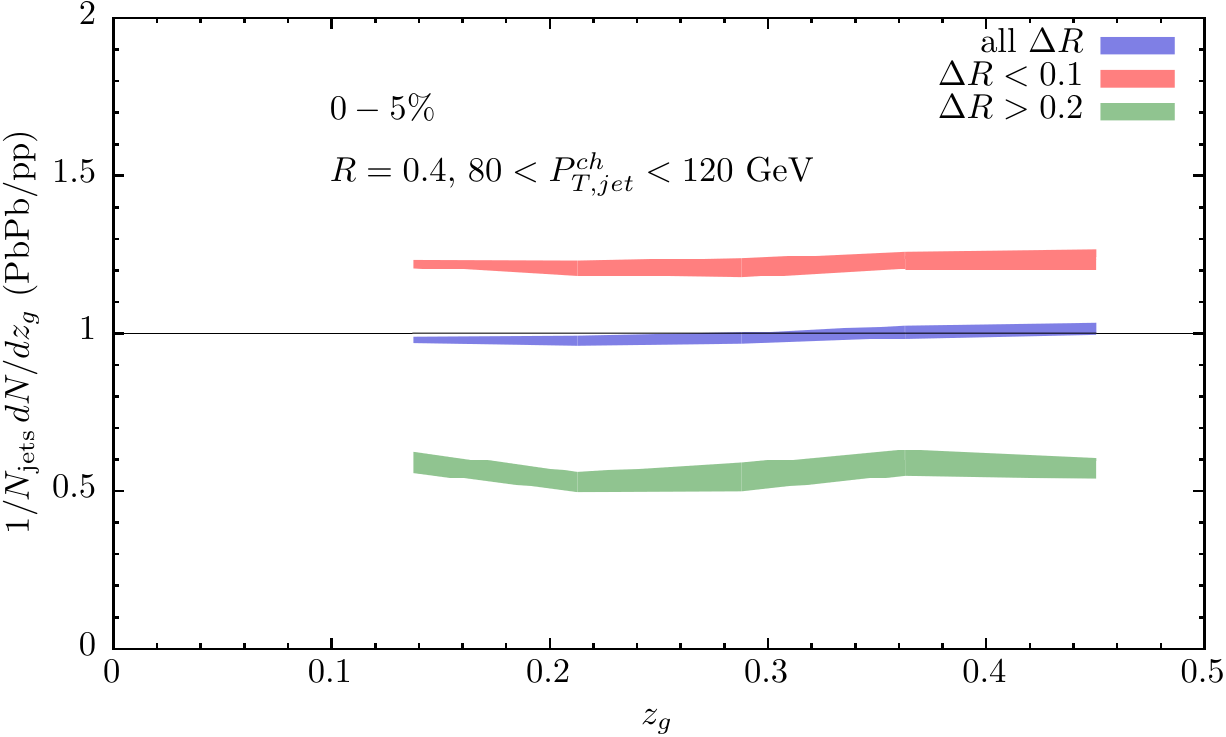}\hfill
	\vspace{-0.1in} 
	\includegraphics[scale=0.65]{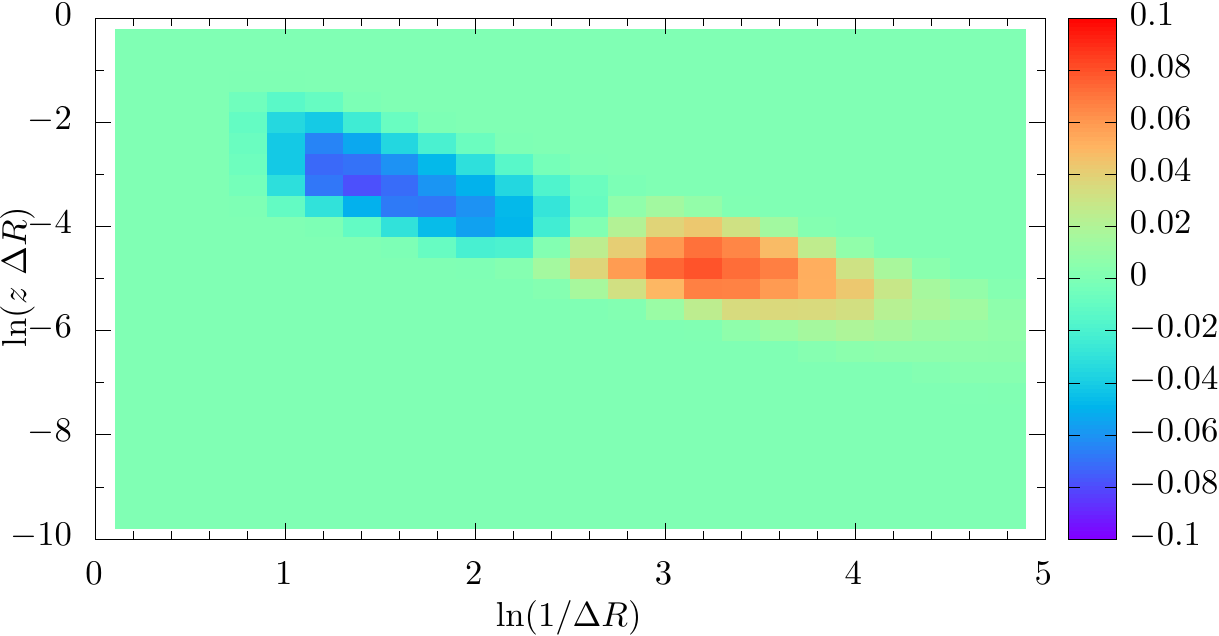}
	\hspace*{\fill}
\vspace{-0.1in}	
	\caption {Left: Ratio of the momentum sharing distribution $z_g$ with respect to vacuum, for different angular cuts $\Delta R$ on the distance between the two branches. All the distributions are normalized to the number of jets that fall within the jet $p_T$ range, $N_{\rm jets}$, following ALICE practice. Right: The difference between the medium and vacuum results for the Lund plane distribution.}
	
	\label{both}
\end{figure}


In the left panel of Fig.~\ref{both}, we focus on results regarding the medium-induced 
modification of the distribution of the 
momentum sharing fraction $z_g$ \cite{Larkoski:2014wba} for the largest angle hard branching, expressed as the ratio of the hybrid model result over the vacuum result.
We have produced three different bands, corresponding to different cuts on the angular separation $\Delta R$ between the two branches that satisfy the Soft Drop condition. We focus first on the blue band,  corresponding to the results inclusive in $\Delta R$. 
The first thing to note is that the ratio is consistent with one.  
This might have been expected, given that our model includes no additional splittings induced by interaction with the medium. However, looking at our results differentially in $\Delta R$ 
shows that in fact quenching has substantial effects, which happen to cancel in the inclusive result.
When we select jets with a branching with $\Delta R<0.1$ between the two branches, as in the red band, we see an enhancement in the quenched sample. This corroborates the observation, extracted from Fig.~\ref{nSD}, that narrow jets are overrepresented in the quenched sample relative to their representation in the vacuum jet ensemble. In contrast, when we select jets with a branching with $\Delta R>0.2$, the green band, we see that these wider jets are underrepresented after quenching.
%
%
Satisfying the Soft Drop condition for the first time with a larger $\Delta R$ indicates that the jet is wider, has a larger jet mass, contains more independent sources of energy loss as seen in Fig.~\ref{both}, and has more available phase space for further splittings as seen in Fig.~\ref{nSD}.
%

Finally, we conclude with results for the primary emission Lund plane distribution, shown in the right panel of Fig.~\ref{both}. This density map represents a very promising tool for a detailed analysis of the medium modification of jet substructure \cite{marta}, as it very conveniently depicts 
the joint distribution of the angular separation $\Delta R$ and the momentum sharing $z$ of the first Soft Drop splitting in a sample of jets.
The right panel of Fig.~\ref{both} shows the difference between this distribution in our hybrid model and vacuum results. It cleanly and clearly shows that quenching serves to enhance the representation of jets containing branchings with lower values of $\Delta R$ (excess of narrow jets) and deplete the representation of jets with branchings with larger $\Delta R$ (depletion of wide jets) which is exactly as we would expect based upon our results in Fig.~\ref{nSD} and the left panel of Fig.~\ref{both}.


\section{Conclusions}

Our results highlight the role that jet-by-jet variations in the angular widths of jets with a given $p_T$, hence in the number of splittings,  hence in the number of independent sources of energy loss, have on jet substructure observables like $z_g$ and $\Delta R$ as well as how the modification of the distributions of these observables 
by quenching arises because quenching enhances the representation of narrower jets with fewer splittings in the ensemble.
The qualitative agreement between our results for the $n_{\rm SD}$ and Lund plane distributions in the hybrid model 
and preliminary data from ALICE \cite{nima} support this picture.

A fair comparison between the modification of the $z_g$ distribution in our calculations and that in data is not possible without smearing our theoretical results or unfolding the experimental data. 
Given that the jets in the quenched and vacuum ensembles are different in shape, we should expect the effects of smearing to differ, meaning that smearing must modify our results for the ratio plotted in the left panel of Fig.~\ref{both}.
Since the quenched jets in the numerator of the ratio are narrower than the vacuum jets 
in the denominator, it is likely that if $\Delta R$ is chosen to be large 
their $z_g$-distribution will be more affected by
the presence of a fluctuating background at the periphery of the jet. 
We therefore expect that smearing should have less of an effect on the ratio of $z_g$ distributions 
for $\Delta R<0.1$
in the left panel of Fig.~\ref{both}, whereas it may significantly influence the shape of this ratio
for $\Delta R>0.2$ (or for $\Delta R>0.1$, as in the CMS analysis \cite{dhanush}). 
In particular, including the effects of smearing should push the green band in this Figure 
up the most at smaller values of $z_g$, near 0.1, where the softer branch is the softest and hence is most affected by background fluctuations.
The quantitative analysis of the effects of smearing on $z_g$, and how they affect the interpretation of this data, will be presented in future work.

{\bf Acknowledgments:} Work supported by  grants SGR-2017-754, FPA2016-76005-C2-1-P and MDM-2014-0367, by Funda\c c\~ao para a Ci\^encia e a Tecnologia (Portugal) contract CERN/FIS-PAR/0022/2017, by US NSF grant ACI-1550300 and by US DOE Office of Nuclear Physics contract DE-SC0011090.

\end{document}